April 29, 2016

# A framework for philosophical biology

Sepehr Ehsani

*Department of Laboratory Medicine and Pathobiology, University of Toronto, Toronto, ON M5S 1A8, Canada*
*Present address: *Computer Science and Artificial Intelligence Laboratory, Massachusetts Institute of Technology, Cambridge, MA  02139, United States*
ehsani@csail.mit.edu

**ABSTRACT**
Advances in biology, at least over the past two centuries, have mostly relied on theories that were subsequently revised, expanded or eventually refuted using experimental and other means. The field of theoretical biology used to primarily provide a basis, similar to theoretical physics in the physical sciences, to rationally examine the frameworks within which biological experiments were carried out and to shed light on overlooked gaps in understanding. Today, however, theoretical biology has generally become synonymous with computational and mathematical biology. This could in part be explained by a relatively recent tendency in which a "data first", rather than a "theory first", approach is preferred. Moreover, generating hypotheses has at times become procedural rather than theoretical, therefore perhaps inadvertently leading some hypotheses to become perfunctory in nature. This situation leaves our understanding enmeshed in data, which should be disentangled from much noise. Given the many unresolved questions in biology and medicine, big and small, ranging from the problem of protein folding to unifying causative frameworks of complex non-Mendelian human diseases, it seems apt to revive the role of pure theory in the biological sciences. This paper, using the current biomedical literature and historical precedents, makes the case for a "philosophical biology" (philbiology), distinct from but quite complementary to philosophy *of* biology (philobiology), which would entail biological investigation through philosophical approaches. Philbiology would thus be a reincarnation of theoretical biology, adopting the true sense of the word "theory" and making use of a rich tradition of serious philosophical approaches in the natural sciences. A philbiological investigation, after clearly defining a given biological problem, would aim to propose a set of empirical questions, along with a class of possible solutions, about that problem. Importantly, whether or not the questions can be tested using current experimental paradigms would be secondary to whether the questions are inherently empirical or not. These issues will be illustrated using a range of specific examples. The final goal of a philbiological investigation would be to develop a theoretical framework that can lead observational and/or interventional experimental studies of the defined problem, a framework that is structured, generative and expandable, and, crucially, one that simplifies some aspect(s) of the said problem.

**MAIN TEXT**
Biology, particularly molecular biology, has experienced significant changes and technical innovations in the past several decades. Whereas novel insights and descriptive modes of understanding began to emerge from the early applications of molecular biology in the 1970s and 1980s, the widespread popularity of high-throughput techniques and genome sequencing in the 1990s and 2000s led to the generation of a tremendous amount of new insights and descriptive data about the workings of the cell under normal and disease conditions [1]. The human impact of these findings has been especially pronounced in the case of monogenic and/or relatively rare Mendelian diseases. Moreover, among human cancers, one can point to successful primary treatments of lymphomas/leukemias (see for example refs. [2,3,4]) and to overall "annual reductions of 1 to 2% in age-adjusted cancer mortality rates in the United States for many years" [5]. In the domain of infectious diseases, the recently rising cure rates in hepatitis C cases are especially noteworthy [6]. However, contrary to initial expectations, most common human diseases have remained refractory to various (non-symptomatic) therapeutic interventions, mostly because we have not been able to unify the diseases under common causative models and/or mechanisms. In fact, we may sometimes observe in molecular biology research that finding a new "mechanism" in a cellular process comes to only mean finding "a molecule that is involved in the process" [7], which is clearly not in the true spirit of what a "mechanism" entails. Furthermore, although vaccines or immunotherapies are exceptionally effective when applied to preventing infections or treating a particular neoplasm, one is in essence redirecting the means of a complex and still-to-be-understood component of a natural system (i.e., the immune system) to treat a disorder in another part of that same system, thereby circumventing a direct engagement with the system. Overall, it is evident that the mechanistic complexity and heterogeneity of common diseases have proven difficult to simplify [8,9]. Although there have been many suggestions to view each common disease merely as an umbrella term for different constituent smaller diseases, there nevertheless appears to be a common theme and collection of phenotypes that unite all manifestations of each common disease. The question, therefore, is: what is preventing us from reaching a real and unifying understanding of these pathologies?

One can posit that the most plausible answer to the question above is a lack of suitable explanatory and predictive theories in today's biology. It is evident that theory has always been an indistinguishable part of the biological sciences, from ecology [10], evolutionary theory and microbiological basis of disease [11] to the elucidation of DNA and protein structures and networks of gene regulation [12,13]. Nevertheless, in the era of high-throughput and big-data experiments, a notion has become prevalent that observations and data collection can be pursued independently of prior theories. This cannot be the case, since no process of data collection, however carefully planned, can be completely devoid of bias [14]. In the words of Albert Einstein, "it is the theory that determines what we can observe," a statement which was followed by Werner Heisenberg's comment that "we have to remember that what we observe is not nature in itself, but nature exposed to our method of questioning" [15]. A pertinent example here could be the utility of Alan Turing's theoretical reaction-diffusion model to the recent observation of "Turing-like features in the periodic pattern of digits" in developing limb buds [16,17]. Presently, theoretical biology, with some exceptions, has become mostly synonymous with computational biology and the application of mathematical models to various forms of data structures (for some examples, see refs. [18,19,20,21,22,23,24]). The time is therefore ripe to reintroduce genuine theoretical analysis back into biology. But where would new theories come from? One source could be philosophy, in the form of philosophical biology (henceforth referred to as philbiology). As a side note, the phrase "philosophical biology", or "philosophical science" in general, would have seemed pleonastic to the scientists of the Enlightenment and later periods, but today this pleonasm may be necessary. Philbiology, distinct from philosophy *of* biology (philobiology), could endeavor to search for, propose and develop questions and answers in the true spirit of the theoretical sciences using a vast array of tried-and-tested analytical philosophical tools that have been developed over many centuries. It has to be emphasized that the goal is not to produce theories only, because, in the words of the evolutionary biologist Richard Lewontin, "there is no end to plausible storytelling" [25]. Rather, the theories should be accompanied by inherently-testable sets of questions and possible solutions. The aim of this paper is to suggest a framework whereby philosophical approaches can find their way back into mainstream biological research, in the form of a new philosophical biology.

### *Current State of Philosophy in Biology*
The paramount goal of biology, as in the other natural sciences, has been to better understand facets of the natural world through simplification and generalizable "rules". For example, the process of simplification has been evident in theoretical physics [26] or the field of biolinguistics over the past several decades. Robert C. Berwick and Noam Chomsky recently note that "complex linguistic rule systems are now a thing of the past; they have been replaced by much simpler, hence more evolutionarily plausible, approaches" [27]. Moreover, they emphasize that "without serious understanding of the fundamental nature of some system, its manifestations will appear to be chaotic, highly variable, and lacking general properties" [27]. Recently and in the past, however, not all goals in the natural sciences have been directed towards simplification per se. The nature of some questions has historically been more metaphysical. For example, a commentary appearing in the *North American Review* in 1868 stated that the "great questions of biology, considered in its philosophical aspect, are three: What is the origin of life in the first instance? What is the origin of species or the different forms of life? What are the causes of organic evolution in general?" [28].

Given that simplification is a form of knowledge, how has "knowledge" been described historically in the natural sciences and philosophy? In his magnum opus *Ethics* published in 1677, Baruch Spinoza draws a distinction among three types of knowledge: (i) "knowledge of the first kind" being *opinion* or *imagination*, (ii) "knowledge of the second kind" being *reason*, and (iii) *intuition* forming "knowledge of the third kind" [29]. Although the definitions and common usages of these terms may differ in various respects today compared to the seventeenth century, one can argue that although *reason* is generally held as the salient type of knowledge in biology and the natural sciences, all three types of knowledge nevertheless contribute to what we consider as our collective understanding of biology. Bertrand Russell differentiated between "knowledge by acquaintance" and "knowledge by description" [30]. Looking at these from another angle, and using the philosophical notion of *Verstehen* that is used in the social sciences to study how one *relates to that which one wants to understand*, we can perhaps **distinguish between two types of understanding: participatory and non-participatory**. Let us begin with the latter type. Non-participatory (or intrinsic) understanding can, for example, refer to the type of understanding that the human mind can have of the primary colors. A person can "understand" and be "knowledgeable" of the colors perfectly and in a plethora of contexts, but cannot begin to truly explain, describe or convey that understanding. In this way, facets of the conscious mind are *not* "participating" in forming that understanding. Non-participatory understanding is thus devoid of explicit cause-and-effect features or forms of measurement, and the mantra of "through measuring to knowing" [31] cannot hold true. It is perhaps suitable to quote the philosopher and psychologist Wilhelm Dilthey who noted in 1894 that "we explain nature, [whereas] we understand psychic life" [32], drawing a distinction between explanation and understanding.

Participatory (or external) understanding can be exemplified with understanding the mechanism of a clock, and the ability to explain and reconstruct it. We can actively and consciously "participate" in forming this type of knowledge. Participatory understanding is typical of the kind of understanding the natural sciences can initially aim for, which may not resonate with internal forms of understanding in the mind and would therefore have to be mechanistic and descriptive [33]. This is a main reason for the widespread use of metaphors in biology. In protein biology, for example, we talk of "folding", "binding affinity" [34], "liquid-to-solid phase transition" [35] and "liquid droplets" [36], to name a few. Representations and diagrams, such as actograms in circadian rhythm research [37,38], are also essentially metaphorical. "Emergence" is another popular metaphor, particularly in the neuroscience field, which is unfortunately an intractable notion [39]. As an aside, emergent behavior in biology was modeled using partial differential equations beginning in the 1970s [40], and later used as a means to explain phenotypes of a genetic reductionist model. However, it may be fair to argue that this and similar notions have not aided in augmenting mechanistic understanding in biology. In fact, the concept of emergence can be traced back to condensed-matter physics and the still-present gap between the understanding of individual components (e.g., electrons) and resultant phenomena (e.g., a material's electronic properties) [41]. Here, in the case of electrons and electronic properties, although the mantra of "the whole being more than [or other than] the sum of its parts" [42] rings true, serious investigations of the connections between the parts and the whole can begin only when it can be claimed with some certainty that *all* the parts are in fact known.

Metaphors are problematic because they borrow from components of our intrinsic understanding to aid in external understanding. Metaphors can certainly be helpful at first, but should eventually be replaced with a suitable explanatory theory. In other words, the aim in fields as diverse as protein folding or neuroscience should be to "define a technical notation in the context of an explanatory theory" [43]. In other words, "approximating observational phenomena is very different from formulating an explanatory account of a significant body of empirical data" [44]. If such explanatory theories are based on sound philosophical reasoning, the results could be fruitful. Lymphocytic V(D)J recombination in adaptive immunity [45] and the elucidation of friction reduction by bacteria in their medium [46,47,48] are two instances of the successful implementation of a strong theoretical model through to experimental validation. Another example of the usage of sound theoretical arguments in advance of establishing an experimental paradigm is the recent focus on siderophore quenching strategies to avoid the emergence of antibiotic resistance in a bacterial community [49], which in essence shift the burden of antibiotic resistance from individual bacterial cells or colonies to a microbial community. In fact, it appears that the immune system utilizes a similar strategy as part of its own defensive mechanisms [50].

***An Abridged Selection of the History of Philosophy in Medicine and Biology***
Building on the few examples provided above, it may be useful to have a more in-depth look at some historical precedents of philosophical reasoning in what we now call the biomedical sciences. To begin with, in 1235 AD, Nasir ad-Din Tusi completed his wide-ranging magnum opus *Akhlaq-i Nasiri* ("The Nasirean Ethics"), in which, among other topics, he postulated one of the earliest known observation-based theories of speciation and adaptation. Furthermore, he categorized the different branches of philosophy, including the natural sciences, as follows (transl. G. M. Wickens, 1964):
> "Knowledge of existent things is […] in two divisions: that relating to the first division, called *Speculative Philosophy*; and that of the second division, called *Practical Philosophy*. Speculative Philosophy itself is in two divisions: a knowledge of that, the existence of which is not conditional on involvement with matter; and, secondly, a knowledge of that which cannot exist so long as there be no involvement with matter. This latter division is also twice divided:

on the one hand is that, into the intellection and conception of which consideration of involvement with matter does not enter as a condition; on the other, is that which is known only by consideration of involvement with matter. Thus, in this way, there are three divisions of Speculative Philosophy: the first is called *Metaphysics*, the second *Mathematics*, and the third **Natural Science**."

This early account acts as a suitable framework to consider later applications of philosophical approaches to understand, in the words of Tusi, the "existent things". In the first volume of his 1689 "An Essay Concerning Humane Understanding", John Locke offered an account of causality that still informs today's use of mechanistic explanations in molecular biology: "For to have the idea of cause and effect, it suffices to consider any simple idea or substance, as beginning to exist, by the operation of some other, without knowing the manner of that operation." Nevertheless, the tension between experimental and theoretical approaches in biology too has historical roots. One can, for example, point to differing modes of explanation in the seventeenth century of "spontaneous generation" in embryology [51], or the eighteenth century Scottish surgeon John Hunter's suggestion to his vaccine-pioneering student Edward Jenner to "Don't think. Try" [52,53]. In the 1850s, Rudolf Virchow, the pioneer of cellular pathology [54], is famous for having quoted Salomon Neumann that "medicine is a social science" [55]. He also stated that "medicine as a social science, as the science of human beings, has the obligation to point out problems and to attempt their theoretical solution; the politician, the practical anthropologist, must find the means for their actual solution" [56].

Here it may be worthwhile to point to some specific cases from the past two centuries where philosophy, theory and experimentation demonstrate an intertwined relationship. In 1806, Theodor Grotthuss proposed a theory of proton tunneling across hydrogen bonds [57]. The Grotthuss mechanism remains an enigmatic and very relevant question and phenomenon in studies of water structure and water-protein interactions. Theoretical investigations into hydrogen bonding in water remain an active area of research, as for example David C. Clary recently notes that "the excellent detailed agreement between the quantum dynamical calculations and experimental data shows that theory is getting much closer to a highly accurate description of water and, thus, to providing a detailed quantitative understanding of hydrogen-bond dynamics" [58]. In 1872, Casimir Davaine put forth the idea of "passages" in microbiology by studying *Bacillus anthracis* virulence in blood samples, a concept that has been a mainstay of any microbiological experiment up to this day. The duality of humoralism versus cellularism began to take shape in 1882, with the development of Ilya Metchnikoff's theory of phagocytosis [59]. Although initially derided by some as a "fairy tale" [60], this theory still resonates today in immunological research on phagocytic cells and inflammation [61]. Around the same period, Louis Pasteur's "discovery that microbes discriminate between *D*- and *L*-substrates […] had been given little attention until taken up by [Emil] Fischer, who suggested that 'the yeast cells with their asymmetrically formed agent are capable of attacking only sugars of which the geometrical form does not differ too widely from that of *D*-glucose'" [62], leading to Fischer's proposition of a "lock and key" metaphor in 1894. The interaction between theory and observation wasn't always harmonious in hindsight. For example, *Bacillus icteroides* was proposed in 1896 (by Giuseppe Sanarelli) as a bacterial cause of yellow fever which fulfilled Koch's postulates. Similarly, based on the prevalent germ theory of the early twentieth century, investigators' finding of the bacterium *Haemophilus influenzae* in influenza patients was perfectly reasonable and a suitable answer for the cause of influenza. Only a few decades later was a virus identified as the cause through the work of Richard Shope and colleagues [63]. In fact, Oswald T. Avery's work on DNA was a result of his decades-long work on influenza and pneumonia in the same period (for an in-depth discussion, see ref. [52]).

The examples above show approaches that started with a philosophical theory followed by selective experimentation, leading eventually to a refined theory. Nevertheless, there were instances where this model did not apply. For example, in 1909, Paul Ehrlich and his collaborators tested nine hundred chemical compounds for a syphilis treatment to eventually identify Salvarsan, which we could venture to call one of the earliest precursors (since the heydays of alchemy) of today's high-throughput compound screens. This approach relied more on trial-and-error than *ab initio* theorizing. One can also point to important biological questions for which theory and heuristics have played less pronounced roles individually, the results of which are still being explored today. For example, since the work of Frederick Banting, Charles Best and Nicolae Paulescu in the early 1920s on insulin extracts, we now have detailed descriptions of insulin signaling cascades using high-throughput phosphoproteomics [64]. Moreover, around 1945, Søren L. Ørskov's biochemical studies of the diffusion of metabolites across the yeast cell membrane led him to compare and contrast a "pore theory" and a "lipid solubility theory" of the diffusion process [62,65]. The roles of protein channels and the phospholipid bilayer in acting as an interface between the intracellular and extracellular environments are as relevant today as seven decades ago.

The first half of the twentieth century saw the formalized emergence of theoretical and experimental branches in physics, a division that might not have seemed necessary beforehand. Biology saw similar developments. In the 1930s, after the work of investigators such as J. B. S. Haldane, Ronald Fisher and Sewall Wright had established modern synthesis in evolutionary biology, Conrad H. Waddington and Ludwig von Bertalanffy, among others, proceeded towards formalizing "theoretical biology" and "systems biology", respectively [66]. In 1968, Marjorie Grene published "Approaches to a Philosophical Biology" on the state and outlook of the philosophy of biology and, over the next several decades, elements of philosophical approaches to biology were further extended into medical humanities (e.g., in Hans Jonas's 1966 publication on bioethics in "The Phenomenon of Life: Toward a Philosophical Biology"), philosophical psychology, philosophy of chemistry, philosophical chemistry (tracing its roots at least back to the work of Joseph Black in the eighteenth century), physical oncology [67], healthcare improvement theory [68] and other related disciplines. Nevertheless, with the advent of recombinant DNA technology in the 1970s followed by molecular biology and the widespread adoption of relevant technologies such as flow cytometry [69], theoretical/philosophical biology did not have an opportunity to reach the same level of attention as its counterpart in physics, and today, as noted earlier, many consider theoretical biology to be synonymous with computational and mathematical biology. This is not to say that questions pertaining to theoretical biology have been forgotten. Some of these questions have indeed been rigorously pursued under the domain of philosophy of science/biology, focusing on problems in evolutionary theory or population genetics, amongst others [70,71].

*Philbiology: Theory and Practice*
Based on these historical precedents, the point of philosophical biology would be to renew the application of philosophical reasoning to theoretical biology research. As noted earlier, philbiology would be distinct from philosophy of biology, which is primarily a philosophical/historical approach to the development of the biological sciences up to the present (it should be noted here that historical analysis may or may not be included in philosophy of biology investigations in different circles). Philbiology is biology through philosophy, while philobiology (in its pure form) is, by definition, philosophy and history through biology. Philbiology aims to gain insights into foundational questions in biology using a philosophical approach. Its objectives would, in essence, be similar to the ideals of the physical sciences community in the early period of theoretical physics in the 1920s and 1930s. In fact, Max Born commented in 1963 that "I am now convinced that theoretical physics is actually philosophy". Nevertheless, although the goals would be similar, it is evident that biology and physics are dissimilar in many ways and not necessarily reducible to each other. As Berwick and Chomsky point out, "biology is more like case law, not Newtonian physics" [27].

A primary concern of philbiology would be on the development of models in biology. Baruch Spinoza pointed out a common fallacy regarding models in his *Ethics* (1677): "For men are wont to form general ideas both of natural phenomena and of artifacts, and these ideas they regard as models, and they believe that Nature […] looks to these ideas and holds them before herself as models. So when they see something occurring in Nature at variance with their preconceived ideal of the thing in question, they believe that Nature has then failed or blundered and has left that thing imperfect" [29]. In other words, testing a model *against* a natural phenomenon is different than testing the said natural phenomenon *against* the said model, an issue which is as relevant today as it was more than 300 years ago. Jeremy Gunawardena defines a model as "some form of symbolic representation of our assumptions about reality" [72], with "assumptions" here being a key word. He further describes the duality between informal models (mental, verbal, etc.) and formal models (mathematical) in biology. **Whereas assumptions about reality are tested daily in the laboratory, and formal models are developed in computational/mathematical biology, informal models, as a bridge between our assumptions about reality and symbolic representations of those assumptions, are ripe for philosophical investigation** [73]. A second primary concern of philbiology would be on studying the "limits" of our current understanding in biology. What is accessible to us today and what is inaccessible? What can we reasonably expect to find and understand about the cell, given that not finding something does not indicate its non-existence [74]? Again drawing on an analogy with theoretical physics, the following observation from the physicist Jean Baptiste Perrin in his 1926 Nobel Lecture [75] is pertinent:

> "Certain scholars considered that since the appearances on our scale were finally the only
> important ones for us, there was no point in seeking what might exist in an inaccessible
> domain. I find it very difficult to understand this point of view since what is inaccessible today may become accessible tomorrow (as has happened by the invention of the microscope), and also because coherent assumptions on what is still invisible may increase our understanding of the visible." [75]

There is at present some biological research that could, on closer inspection, be categorized as falling under the philbiology category. The work of Mariscal and Doolittle on the origins of eukaryotes [76] or that of Wei, Prentice and Balasubramanian on using mathematical modeling to propose that a "principle of economy predicts the functional architecture of grid cells" [77], both in 2015, are two such studies. Nevertheless, it would be prudent here to suggest a few specific examples of questions and topics that a philosophical biologist could consider (**Figure 1**). These examples are divided into two categories: Theoretical Methods and Tools (TMT) and Theoretical Problems and Solutions (TPS):

**TMT**
When we mention "philosophical methods", the logical operations of *deduction* and *induction* are usually invoked. To be specific, using the case of the cellular basis of time [78] as an example, we could say the following: (A) If (i) *all biological reactions in cellular environments are synchronous* and (ii) *all synchronous phenomena need an internal or external pacemaker*, then we can syllogistically deduce that (iii) *all biological reactions in cellular environments are definitely driven by a pacemaker*. (B) If (i) *cell line A has a pacemaker* and (ii) *cell line B also has a pacemaker*, then we can induce that (iii) *other cells may also have a pacemaker*. These two methods are very useful and are in fact indistinguishable components of human rational reasoning in general. Nevertheless, in proposing a set of methods and tools for philbiology, we could develop approaches that are more specific and tailored to the kinds of questions that are investigated.

The Theoretical Methods and Tools (TMT) category can encompass various utilizations and developments of philosophical (and related) approaches for applications to specific biological questions (see for example ref. [79]). These approaches could be analytical, following the works of philosophers such as Gottlob Frege, Alfred Tarski and W. V. Quine, or could follow non-analytical and non-traditional reasoning methods. If a philbiological investigation foresees a direct or an immediate human impact, the philosophical approach should be grounded in moral philosophy. As much as possible, one could aim to initially avoid using philosophical methods that provoke competing or non-trivial definitions (e.g., mereological, teleological, epistemological, tautological, phenomenological, ontological, normative, etc.) and to appeal to as-simple-as-possible rational and common-sense approaches. Nevertheless, certain "simplified" components of the concepts in the former category should necessarily be used. Some examples include:

1. When we attempt to understand and describe the behavior of a protein or lipid membrane in a cell, how do we begin to offer a "good" explanation? Is a molecular "descriptive" account an "explanation" nonetheless? Here contemporary analytical philosophical methods that have been developed at least beginning with the work of Rudolf Carnap (e.g., with regards to "explication") can have great utility [80,81].

2. In offering a descriptive or causal explanation, how do we move beyond providing a statistical view of the phenomenon at hand [82]? Given current trends in biology and the natural sciences in general towards the expansion of numerical models and big-data science [83], this question becomes especially important, as no natural process can have a "statistical nature"; a natural process just has a "nature", which we may choose to model statistically in the absence of a suitable explanatory theory. In fact, some historians of science rightly point out the fact that "big data" is not a new notion in the sciences, as large collections of data have been a staple of astronomy, in the form of astronomical tables, for many centuries [84,85].

3. Because many biological interactions happens at infinitesimal scales where exact measurements give way to approximations, can "non-standard analysis" and the theory of infinitesimals (developed by Abraham Robinson; published in 1966) along with hyperreal numbers [86] be used instead of standard calculus? Can this be combined with Gödel numbering, mereology and set theory?

4. In trying to establish causal relationships in gene/protein circuits, how can we use deontic logic (which focuses on the notion of "obligation")? Would deontic logical approaches help with questions such as *does something that looks like an "effect" really need a "cause"*? What role could non-classical logic play? Here one should note that although deontic and non-classical logical systems are themselves divided into various subcomponents, the applicability to philbiology would not necessarily be in the closed formal proofs that these systems allow, but more in the processes and connections that they can hint at or disprove.

**TPS**

The Theoretical Problems and Solutions (TPS) category comprises a group of new questions and specific sets of possible solutions that are proposed using philosophical approaches. Moreover, the goal may at times be to come up with new or improved theories. A "good" theory is an adaptable theory, one that would allow for incremental growth in understanding while also hinting at gaps that a new and improved theory can fill [87] and, in a sense, produce a leap in understanding. Furthermore, such a theory should be developed in "abstraction from the full complexity" [88,89] of what is being studied. TPS also includes existing questions or paradigms that are expanded or modified. Some examples include:

1. What is the difference between two helical or beta-sheet domains of equal length in two proteins arising from different amino acid sequences? Is a disordered domain of a protein really "disordered", or do such domains adopt a limited set of structures that are "appropriate to, but not caused by" [90] the protein/lipid microenvironment around the protein? Can new protein folding theories become alternatives to molecular dynamics simulations [91]? Is it conceivable that in some circumstances protein folding, rather than proceeding to minimize free energy [92], proceeds primarily to minimize search efficiency only (i.e., "computational" efficiency from the perspective of the amino acid sequence)? These questions are not only important in understanding the structure and dynamics of proteins, but are also indispensable in deriving new theories to account for protein aggregation in neurodegenerative diseases or prion-like propagation of proteins such as the tau protein [93,94]. Furthermore, these questions all have unresolved theoretical underpinnings that can be resolved piece by piece using philosophy. For example, if one tries to conceptualize the number of possible atom-to-atom "interactions" (a vague notion that needs resolution itself) as a nascent polypeptide chain emerges from the ribosome, the number of possibilities can easily escape finite bounds, whereas it is evident that protein folding takes place in a finite amount of time in the cell or even in artificial conditions. An infinite number of possibilities resolving in a finite amount of time is reminiscent of a "supertask" in mechanical philosophy [95], a concept that has been explored since the Antiquity.

2. Given the many ambiguities about the structure of water molecules [96], how do hydrophilic residues of proteins really interact with water molecules in their vicinity? Are such interactions always electrostatic in nature, or could non-electrostatic interactions, such as hydrogen-hydrogen (H-H) bonding (which is distinct from electrostatic hydrogen bonding [97]), also play an important role? It should be noted that water is by no means the only "simple" ubiquitous molecule for which deep ambiguities remain. The C-H bonds of the seemingly simple methane molecule ($CH_4$) are another case in point [98], a strand of investigation which could have implications for C-H bonds in amino acids. Moreover, given the essential interaction of many proteins with metal ions, organometallic compounds and other small molecules, do current theories satisfactorily account for the unique interaction of amino acids and these compounds? (As an example, see ref. [99] for a discussion of the challenges in understanding the structure of the organometallic compound ferrocene.) It should be noted that these questions are essential *primary questions* not only to understand protein folding, but also protein interactions with other proteins and macromolecules. Before finding suitable answers to these questions, it is doubtful that an explanatory framework can ever reach a point so that broader concepts, such as cross-species "inter-interactomes" of protein-protein interaction networks [100], could be addressed.

3. What is the concept of time in the cell (cellular time as opposed to circadian time) [78]? Does a cell need a sub-second timekeeping or pacemaking mechanism to arrange the plethora of simultaneous functions taking place in the cytoplasm and other subcellular compartments? If so, what can such a mechanism be?

4. Can one deduce whether the timeframe of a given cellular process increases polynomially with the increasing complexity of the task? This problem could borrow from extensive research in theoretical computer science under the famous *P* (polynomial) versus *NP* (nondeterministic polynomial) paradigm. Such questions can initially draw from existing research in computational biology regarding the protein folding problem [101] or RNA structure prediction [102], to name a few. One can eventually expand the scope of these investigations to include kinetic studies of enzymes and process timing in the framework of the Michaelis-Menten equation [103].

5. What exactly is aging on an organism level, and what accounts for the diversity of aging profiles across species [104,105] and different phyla [106,107]? As evidenced by work on the *Prochlorococcus* genus [108], this line of investigation would require an analysis of what it really means to be a species, a genus, etc. [109,110].

6. For the phenomenon of antimicrobial resistance, as alluded to earlier, can one devise a strategy where the emergence of resistance would theoretically be impossible? Since the introduction of sulfonamide antibacterial drugs in the 1930s, the emergence of resistant subpopulations of bacteria or fungi has become inevitable [111]. Nevertheless, notwithstanding the re-emerging bacteriophage research field [112], strategies are beginning to be refurbished or newly devised where compound-based antimicrobial resistance would become avoidable [113,114,115,116,117]. This is a critical and ripe area for practical philosophical contributions.

7. In cancer biology, what are the theoretical underpinnings of the occasionally paradoxical nature of cell proliferation, metastasis [118], heterogeneous origins [119], differing outcomes [120] and spontaneous regression [121,122,123]? How can some of these paradoxes be used as "natural experiments" [124,125,126] in cancer research? Since the postulation of the Warburg effect more than 90 years ago, our understanding of cancer metabolism, and oncology in general, has greatly advanced. Nevertheless, an all-encompassing theory is still lacking. For example, in reviewing a recent metabolomic analysis of cancer cell proliferation [127], Tanner and Rutter pose the following questions that need explanation [128] (quoted here with permission):

"Although the majority of cell protein is comprised of amino acids imported from the environment, why do cultured cells—awash in amino-acid-rich culture medium—utilize glutamine to synthesize other amino acids de novo even when those amino acids are available for import? Might this have to do with limited import capacity, or is there a separate unforeseen advantage to biosynthesis? Finally, the finding that glucose-derived carbon contributes to a small fraction of cell mass raises still more questions. Why don't proliferating cells utilize the large amounts of carbon consumed as glucose to meet their biosynthetic needs? What is the purpose of such a carbon-wasting metabolic program? Is it simply that this program enables the rapid production of adequate ATP while maintaining the NAD/NADH redox balance, or is there more to it?" [128]

One could posit that these and other questions will only yield to new experimental studies if a new explanatory theory is provided to frame the plethora of pieces of knowledge that are already known in this field. In fact, certain areas within the cancer research field (e.g., cancer stem cells or immunotherapy) have already benefited noticeably from philosophical and theoretical approaches [129,130,131,132].

8. More broadly, what are the inherent differences between correlation and causation [133,134]? Do they have structural differences? For what questions might causal thinking and the notion of agency not be necessary? In areas such as neural networks, to paraphrase the pre-Socratic philosopher Heraclitus of Ephesus, "is a hidden connection stronger than an obvious connection"? Could a hidden or unobservable variable in an experiment, to borrow from economic theory, "share covariance properties with the observed variables" [135]? (Refs. [136,137,138] provide further discussion related to this topic.) Additionally, similar to the notion mentioned earlier for disordered domains in proteins, are there biological processes that are "appropriate to but not caused by" [90] the stimuli that are currently thought to be the causes of those processes? These questions are fundamental to all areas of biology, from investigating the still-unraveling workings of organelles [139,140] to metabolic processes [141], cell-death pathways [142], designing protocols that allow a smoother transition of findings from model organisms to humans [143], and human speech fluency [144], to name a few. It is also evident that a more thorough understanding of causative structures has direct applicability to research on the pathomechanism of diseases regardless of whether the exact etiology is more or less known (e.g., Mendelian [145,146] or infectious diseases [147]) or unknown (e.g., many chronic/complex diseases). A case in point is the ongoing discussion on where the exact cause of neurotoxicity and pathology in Alzheimer's disease truly resides [148,149].

9. Do cellular processes that seem chaotic, stochastic or random [150,151,152,153], such as bursts of transcription or Brownian-like motion of different macromolecules in the cytoplasm, in fact follow as-yet unrecognized deterministic pathways [154,155]?

10. What exactly is uncertainty, and is it possible to postulate theories that go beyond a statistical description of uncertainty? For example, the American Statistical Association has recently emphasized that a *p*-value is "a statement about data in relation to a specified hypothetical explanation, and is not a statement about the explanation itself" [156]. In addition, are there concepts that "ought to be true" but that we cannot describe or observe with any certainty? Are there aspects of biological cells which one can never be certain about? In other words, are there limits to our understanding in certain areas of biology (e.g., see ref. [25])? The "low handing fruits" amongst these questions may initially be found in computational biology. For example, there have been efforts to identify inherent upper limits in accelerating search speeds in biological datasets [157]. Moreover, notions of "loose and tight" limits have been defined for computational problems [158]. Furthermore, similar questions can be asked in chemistry, as in the words of Christopher T. Walsh: "how much new chemistry is yet to be found [and] what kinds of biosynthetic enzymatic transformations are yet to be characterized?" [159].

11. The field of neuroscience is readily conducive to philosophical and theoretical inquiries (e.g., see refs. [160,161,162,163]). However, in light of the numerous unsolved, lower-hanging-fruit problems in "simpler" organisms such as *D. melanogaster* or *C. elegans* [164], many questions regarding human cognition, the primate nervous system or the mouse brain (the circuity of which is beginning to be mapped [165]) may remain outside the purview of philbiology for some time to come. Nevertheless, there are questions that could be further refined in human cognition and neuroscience using philbiology. For example, why are "our brains […] preprogrammed to misread certain images"? [166] or what is the initiating mechanism of voluntary movements [167]?

12. Lastly, certain philosophy of biology threads could be investigated from a philbiological perspective with immediate application to both the philosophy of biology field and philosophical biology. In fact, the *Stanford Encyclopedia of Philosophy* notes that "when addressing [conceptual puzzles within biology], there is no clear distinction between philosophy of biology and theoretical biology". For instance, if chemistry is arguably not reducible to physics (yet unifiable at the same time) [90], in the same spirit could we ask if the properties of a biological system (e.g., a cell) can ever be reduced to the properties of its components (e.g., proteins)? As another example, if one excludes some obvious explanations, why are certain findings either in molecular biology or clinical medicine not reproducible [168]? Approaches to tackle these and other questions of this kind are well-established in the philosophy of science literature, and therefore philbiology and philosophy of science are not exclusive of each other.

### *Philbiology: Inherent and Experimental Verifiability*

Given the various lines of study suggested above in the TMT and TPS sections, what would be an endpoint to, or natural progression of, a philbiological investigation? Is empirical verification a necessary touchstone? Although certain outcomes of philbiological studies can and should be tested computationally or in a molecular biology laboratory, one can posit that experimental validation should not be the ultimate standard to validate or invalidate such an investigation. Again to draw an analogy with theoretical physics, early quantum physicists realized that some theoretical paradigms will, at least for the foreseeable future, remain outside the purview of experimental falsifiability in light of the inherent limits within physical experimental approaches [169]. The recently announced Laser Interferometer Gravitational-Wave Observatory (LIGO) experimental results act as a case in point [170]. The non-obligatory interaction between theory and experiment is not limited to physics but can also be observed in, for example, economics [171] or biolinguistics [172]. It is therefore to be anticipated that philbiology would also generate hypotheses or questions that cannot be tested in the laboratory immediately, but their value would only be shown in time and in perhaps not-so-predictable manners.

Theories developed using philbiology are certainly not endpoints in a given theoretical investigation. They are merely stepping stones toward more complete frameworks and programs. A relevant example here is the culmination of many biological theories and empirical validation attempts that now form the exhaustive set of transcriptional and translational programs in developmental biology [173]. This process can also be observed in biolinguistics, whereby linguistic theories that led to our current state of understanding of Universal Grammar were included in a Minimalist Program, which was then followed amongst other things by the development of the Strong Minimalist Thesis (SMT). SMT is now a robust explanatory framework that can allow linguists to discover the extent to which one can "account for the relevant phenomena of language" [174].

**CONCLUSIONS**
For the modern biological disciplines to produce genuine instances of understanding of the workings of the cell, philosophy must regain its rightful place in the theoretical foundations of biology. This is in line with the development of the natural sciences at least since the Enlightenment. The overarching aim of philbiology could be to define suitably innovative and worthwhile horizons for individual parts of biomedical research, horizons that are not mere pedantic extrapolations of current technical information. Furthermore, solutions that arise from these investigations may be isomorphic, such that their theoretical structure could be applicable to other areas of the sciences, in the same line that a number of frameworks from modern linguistics have been applied in this manuscript in the context of philosophical biology.

It may be apt to end on a note that the complexities and "deep truths" of cellular processes could remain hidden even in spite of philbiology and other sincere efforts. In other words, "the sea will be the sea, whatever the drop's philosophy" (Attar of Nishapur). Nevertheless, one can at least be assured that a philosophical approach to biology will constantly question our questions and provide a framework for reassessing and improving our perspectives of the workings of the cell in normal and disease biology.

**ACKNOWLEDGEMENTS**
I would like to thank the Canadian Institutes of Health Research for fellowship support, and colleagues at the University of Toronto and MIT for helpful discussions. This work is dedicated to the memory of Azadeh Aghamir and the early theoretical researchers of neonatal and pediatric conditions.

**FIGURE LEGEND**

**Figure 1. A proposed outline for philosophical biology investigations.** Philbiology can be framed as a set of perspectives to approach what is known and not known about a given topic in biology. These perspectives could be cognisant of (**1**) analytical, cognitive and rational philosophical reasoning, (**2**) a general goal toward simplification and parsimony, (**3**) novel mathematical, logical or other means of measurement, and (**4**) a general aura of uncertainty around the interface between our cognitive capacity and different hard facets of nature. An investigation that bears philbiology into account can (**A**) use these perspectives and choose one or more philosophical tools to (**B**) approach the problem at hand, using those tools to refine, redefine or even dismiss the initial question. If the question is not dismissed, (**C**) a set of possible solutions could be proposed. The set of possible solutions could eventually be amalgamated into a new theory, which (**D**) may or may not be verifiable based on the current experimental paradigms of the period. (**E**) This process is repeated as more is observed or realized about the said topic. Cell illustration adapted from ref. [78].

# REFERENCES

1. Ehsani S (2013) Macro-trends in research on the central dogma of molecular biology. arXiv:13012397.
2. Schaapveld M, Aleman BM, van Eggermond AM, Janus CP, Krol AD, et al. (2015) Second Cancer Risk Up to 40 Years after Treatment for Hodgkin's Lymphoma. N Engl J Med 373: 2499-2511.
3. Radford J, Longo DL (2015) Second Cancers after Treatment for Hodgkin's Lymphoma--Continuing Cause for Concern. N Engl J Med 373: 2572-2573.
4. Lieberman PM (2014) Virology. Epstein-Barr virus turns 50. Science 343: 1323-1325.
5. Varmus H (2016) The transformation of oncology. Science 352: 123.
6. Rehermann B (2016) HCV in 2015: Advances in hepatitis C research and treatment. Nat Rev Gastroenterol Hepatol 13: 70-72.
7. Garfinkel A (2015) Bad Philosophical Ideas that are Driving Modern Biology and Medicine. MIT Philosophy Colloquia.
8. Wang Y, Waters J, Leung ML, Unruh A, Roh W, et al. (2014) Clonal evolution in breast cancer revealed by single nucleus genome sequencing. Nature 512: 155-160.
9. Ramaswamy V, Taylor MD (2015) Pediatric cancer genomics, a play rather than a portrait. Nat Genet 47: 851-852.
10. Odenbaugh J (2013) Searching for patterns, hunting for causes - Robert MacArthur, the mathematical naturalist. In: Harman O, Dietrich MR, editors. Outsider Scientists: Routes to Innovation in Biology: University of Chicago Press. pp. 181-198.
11. Shou W, Bergstrom CT, Chakraborty AK, Skinner FK (2015) Theory, models and biology. Elife 4: e07158.
12. Britten RJ, Davidson EH (1969) Gene regulation for higher cells: a theory. Science 165: 349-357.
13. O'Malley BW (2010) Masters of the genome. Nat Rev Mol Cell Biol 11: 311.
14. MacCoun R, Perlmutter S (2015) Blind analysis: Hide results to seek the truth. Nature 526: 187-189.
15. Bodner GM (1986) Constructivism: A theory of knowledge. J Chem Educ 63: 873-878.
16. Zuniga A, Zeller R (2014) Development. In Turing's hands--the making of digits. Science 345: 516-517.
17. Raspopovic J, Marcon L, Russo L, Sharpe J (2014) Modeling digits. Digit patterning is controlled by a Bmp-Sox9-Wnt Turing network modulated by morphogen gradients. Science 345: 566-570.
18. Weiss JN, Qu Z, Garfinkel A (2003) Understanding biological complexity: lessons from the past. FASEB J 17: 1-6.
19. Leek JT, Scharpf RB, Bravo HC, Simcha D, Langmead B, et al. (2010) Tackling the widespread and critical impact of batch effects in high-throughput data. Nat Rev Genet 11: 733-739.
20. (2016) So long to the silos. Nat Biotechnol 34: 357.
21. Tadrist L, Darbois-Texier B (2016) Are leaves optimally designed for self-support? An investigation on giant monocots. J Theor Biol 396: 125-131.
22. Asatryan AD, Komarova NL (2016) Evolution of genetic instability in heterogeneous tumors. J Theor Biol 396: 1-12.
23. Armiento A, Doumic M, Moireau P, Rezaei H (2016) Estimation from Moments Measurements for Amyloid Depolymerisation. J Theor Biol 397: 68-88.
24. Bertsch M, Franchi B, Marcello N, Tesi MC, Tosin A (2016) Alzheimer's disease: a mathematical model for onset and progression. Math Med Biol.
25. Lewontin RC (1998) The evolution of cognition: Questions we will never answer. In: Scarborough D, Sternberg S, Osherson DN, editors. An Invitation to Cognitive Science: Methods, Models, and Conceptual Issues (vol 4): MIT Press. pp. 107-132.
26. De las Cuevas G, Cubitt TS (2016) Simple universal models capture all classical spin physics. Science 351: 1180-1183.
27. Berwick RC, Chomsky N (2015) Why Only Us: Language and Evolution: MIT Press.
28. Abbot FE (1868) Philosophical Biology. The North American Review 107: 377-422.
29. Spinoza B (1992) Ethics: with The Treatise on the Emendation of the Intellect and Selected Letters; Feldman S, Shirley S, editors: Hackett Publishing Company.
30. Russell B (1910-1911) Knowledge by acquaintance and knowledge by description Proceedings of the Aristotelian Society 11: 108-128.
31. (2016) The art of measurement. Nat Phys 12: 1.
32. Dilthey W (1977) Ideas Concerning a Descriptive and Analytic Psychology (1894). Descriptive Psychology and Historical Understanding (tr Zaner, RM and Heiges, KL): Martinus Nijhoff Publishers. pp. 21-120.
33. Insel N, Frankland PW (2015) Mechanism, function, and computation in neural systems. Behav Processes 117: 4-11.
34. Vangone A, Bonvin AM (2015) Contacts-based prediction of binding affinity in protein-protein complexes. Elife 4: e07454.
35. Patel A, Lee HO, Jawerth L, Maharana S, Jahnel M, et al. (2015) A Liquid-to-Solid Phase Transition of the ALS Protein FUS Accelerated by Disease Mutation. Cell 162: 1066-1077.
36. Murakami T, Qamar S, Lin JQ, Schierle GS, Rees E, et al. (2015) ALS/FTD Mutation-Induced Phase Transition of FUS Liquid Droplets and Reversible Hydrogels into Irreversible Hydrogels Impairs RNP Granule Function. Neuron 88: 678-690.
37. Jud C, Schmutz I, Hampp G, Oster H, Albrecht U (2005) A guideline for analyzing circadian wheel-running behavior in rodents under different lighting conditions. Biol Proced Online 7: 101-116.
38. Bechtel W, Burnston D, Sheredos B, Abrahamsen A (2014) Representing time in scientific diagrams. Proceeding of the 36th Annual Conference of the Cognitive Science Society. Austin, TX.
39. Brenner S (2010) Sequences and consequences. Philos Trans R Soc Lond B Biol Sci 365: 207-212.
40. Keller EF, Segel LA (1970) Initiation of slime mold aggregation viewed as an instability. J Theor Biol 26: 399-415.
41. (2016) The rise of quantum materials. Nat Phys 12: 105.

42. Upton J, Janeka I, Ferraro N (2014) The whole is more than the sum of its parts: Aristotle, metaphysical. J Craniofac Surg 25: 59-63.
43. Chomsky N (2003) Stony Brook Interviews on the Philosophy of Mind.
44. Everaert MB, Huybregts MA, Chomsky N, Berwick RC, Bolhuis JJ (2015) Structures, Not Strings: Linguistics as Part of the Cognitive Sciences. Trends Cogn Sci 19: 729-743.
45. Dong J, Panchakshari RA, Zhang T, Zhang Y, Hu J, et al. (2015) Orientation-specific joining of AID-initiated DNA breaks promotes antibody class switching. Nature 525: 134-139.
46. Hatwalne Y, Ramaswamy S, Rao M, Simha RA (2004) Rheology of active-particle suspensions. Phys Rev Lett 92: 118101.
47. Lopez HM, Gachelin J, Douarche C, Auradou H, Clement E (2015) Turning Bacteria Suspensions into Superfluids. Phys Rev Lett 115: 028301.
48. Marchetti MC (2015) Soft matter: Frictionless fluids from bacterial teamwork. Nature 525: 37-39.
49. Ross-Gillespie A, Weigert M, Brown SP, Kummerli R (2014) Gallium-mediated siderophore quenching as an evolutionarily robust antibacterial treatment. Evol Med Public Health 2014: 18-29.
50. Nakashige TG, Zhang B, Krebs C, Nolan EM (2015) Human calprotectin is an iron-sequestering host-defense protein. Nat Chem Biol 11: 765-771.
51. Mendelsohn EI (1976) Philosophical biology vs experimental biology: Spontaneous generation in the seventeenth century. Topics in the Philosophy of Biology - Boston Studies in the Philosophy of Science. pp. 37-65.
52. Barry JM (2005) The Great Influenza: The Story of the Deadliest Pandemic in History: Penguin.
53. Bartley S (1999) John Hunter--the scientific surgeon "don't think, try; be patient, be accurate...". J Invest Surg 12: 305-306.
54. Bagot CN, Arya R (2008) Virchow and his triad: a question of attribution. Br J Haematol 143: 180-190.
55. Kottke TE (2011) Medicine is a social science in its very bone and marrow. Mayo Clin Proc 86: 930-932.
56. JRA (2006) Virchow misquoted, part-quoted, and the real McCoy. J Epidemiol Community Health 60: 671.
57. Cukierman S (2006) Et tu, Grotthuss! and other unfinished stories. Biochim Biophys Acta 1757: 876-885.
58. Clary DC (2016) Quantum dynamics in the smallest water droplet. Science 351: 1267-1268.
59. Tauber AI (2003) Metchnikoff and the phagocytosis theory. Nat Rev Mol Cell Biol 4: 897-901.
60. Vikhanski L (2016) Immunity: How Elie Metchnikoff Changed the Course of Modern Medicine: Chicago Review Press.
61. Gordon S (2016) Phagocytosis: An Immunobiologic Process. Immunity 44: 463-475.
62. Barnett JA, Barnett L (2011) Yeast research: A historical overview. ASM Press.
63. Van Epps HL (2006) Influenza: exposing the true killer. J Exp Med 203: 803.
64. Humphrey SJ, Azimifar SB, Mann M (2015) High-throughput phosphoproteomics reveals in vivo insulin signaling dynamics. Nat Biotechnol 33: 990-995.
65. Eddy AA, Barnett JA (2007) A history of research on yeasts. The study of solute transport: the first 90 years, simple and facilitated diffusion. Yeast 24: 1023-1059.
66. Moya A (2015) General Systems Theory and Systems Biology. The Calculus of Life: Springer. pp. 25-30.
67. Frieboes HB, Chaplain MA, Thompson AM, Bearer EL, Lowengrub JS, et al. (2011) Physical oncology: a bench-to-bedside quantitative and predictive approach. Cancer Res 71: 298-302.
68. Davidoff F, Dixon-Woods M, Leviton L, Michie S (2015) Demystifying theory and its use in improvement. BMJ Qual Saf 24: 228-238.
69. Robinson JP, Roederer M (2015) Flow cytometry strikes gold. Science 350: 739-740.
70. Gare A (2008) Approaches to the question, 'what is life?': Reconciling theoretical biology with philosophical biology. Cosmos and History: The Journal of Natural and Social Philosophy 4: 53-77.
71. Gouvêa DY (2015) Explanation and the Evolutionary First Law(s). Phil Sci 82: 363-382.
72. Gunawardena J (2014) Beware the tail that wags the dog: informal and formal models in biology. Mol Biol Cell 25: 3441-3444.
73. Orzack SH (2012) The philosophy of modelling or does the philosophy of biology have any use? Philos Trans R Soc Lond B Biol Sci 367: 170-180.
74. Pass R, Frudd K, Barnett JP, Blindauer CA, Brown DR (2015) Prion infection in cells is abolished by a mutated manganese transporter but shows no relation to zinc. Mol Cell Neurosci 68: 186-193.
75. Perrin JB (1926) Discontinuous Structure of Matter. Nobel Lecture.
76. Mariscal C, Doolittle WF (2015) Eukaryotes first: how could that be? Phil Trans R Soc B 370: 20140322.
77. Wei X, Prentice J, Balasubramanian V (2015) A principle of economy predicts the functional architecture of grid cells. Elife 4: e08362.
78. Ehsani S (2012) Time in the cell: a plausible role for the plasma membrane. arXiv:12100168.
79. Nesse RM (2013) Tinbergen's four questions, organized: a response to Bateson and Laland. Trends Ecol Evol 28: 681-682.
80. Friedman M (1974) Explanation and scientific understanding. J Phil 71: 5-19.
81. Weber E, Van Bouwel J, De Vreese L (2013) How to study scientific explanation? In: ibid, editor. Scientific Explanation: Springer.
82. Lewontin RC (2006) Commentary: statistical analysis or biological analysis as tools for understanding biological causes. Int J Epidemiol 35: 536-537.
83. Bauer P, Thorpe A, Brunet G (2015) The quiet revolution of numerical weather prediction. Nature 525: 47-55.
84. Toomer GJ (1968) A Survey of the Toledan Tables. Osiris 15: 5-174.

85. Mozaffari SM (2016) Planetary latitudes in medieval Islamic astronomy: an analysis of the non-Ptolemaic latitude parameter values in the Maragha and Samarqand astronomical traditions. Arch Hist Exact Sci: 1-29.
86. Robert AM (2011) Nonstandard Analysis: Dover Publications.
87. Brigandt I (2015) Do we need a 'theory' of development? Biol Philos.
88. Martin RM (1980) Primordiality, Science, and Value: State University of New York Press.
89. Chomsky N (1986) Knowledge of Language: Its Nature, Origins, and Use: Praeger.
90. Chomsky N (2009) The mysteries of nature: How deeply hidden? J Phil 106: 167-200.
91. Chung HS, Piana-Agostinetti S, Shaw DE, Eaton WA (2015) Structural origin of slow diffusion in protein folding. Science 349: 1504-1510.
92. Neupane K, Manuel AP, Woodside MT (2016) Protein folding trajectories can be described quantitatively by one-dimensional diffusion over measured energy landscapes. Nat Phys.
93. Wegmann S, Maury EA, Kirk MJ, Saqran L, Roe A, et al. (2015) Removing endogenous tau does not prevent tau propagation yet reduces its neurotoxicity. EMBO J 34: 3028-3041.
94. Walsh DM, Selkoe DJ (2016) A critical appraisal of the pathogenic protein spread hypothesis of neurodegeneration. Nat Rev Neurosci 17: 251-260.
95. Manchak J, Roberts BW (2016) Supertasks. In: Zalta EN, editor. The Stanford Encyclopedia of Philosophy.
96. Thamer M, De Marco L, Ramasesha K, Mandal A, Tokmakoff A (2015) Ultrafast 2D IR spectroscopy of the excess proton in liquid water. Science 350: 78-82.
97. Matta CF (2006) Hydrogen-hydrogen bonding: The non-electrostatic limit of closed-shell interaction between two hydrogen atoms. A critical review. In: Grabowski SJ, editor. Hydrogen Bonding - New Insights: Springer.
98. Smith KT, Berritt S, Gonzalez-Moreiras M, Ahn S, Smith MR, 3rd, et al. (2016) Catalytic borylation of methane. Science 351: 1424-1427.
99. Seeman JI, Cantrill S (2016) Wrong but seminal. Nat Chem 8: 193-200.
100. Zhong Q, Pevzner SJ, Hao T, Wang Y, Mosca R, et al. (2016) An inter-species protein-protein interaction network across vast evolutionary distance. Mol Syst Biol 12: 865.
101. Berger B, Leighton T (1998) Protein folding in the hydrophobic-hydrophilic (HP) model is NP-complete. J Comput Biol 5: 27-40.
102. Smit S, Rother K, Heringa J, Knight R (2008) From knotted to nested RNA structures: a variety of computational methods for pseudoknot removal. RNA 14: 410-416.
103. Xie XS (2013) Biochemistry. Enzyme kinetics, past and present. Science 342: 1457-1459.
104. Jones OR, Scheuerlein A, Salguero-Gomez R, Camarda CG, Schaible R, et al. (2014) Diversity of ageing across the tree of life. Nature 505: 169-173.
105. McCormick MA, Delaney JR, Tsuchiya M, Tsuchiyama S, Shemorry A, et al. (2015) A Comprehensive Analysis of Replicative Lifespan in 4,698 Single-Gene Deletion Strains Uncovers Conserved Mechanisms of Aging. Cell Metab 22: 895-906.
106. Levin M, Anavy L, Cole AG, Winter E, Mostov N, et al. (2016) The mid-developmental transition and the evolution of animal body plans. Nature 531: 637-641.
107. Hug LA, Baker BJ, Anantharaman K, Brown CT, Probst AJ, et al. (2016) A new view of the tree of life. Nat Microbiol 1: 16048.
108. Biller SJ, Berube PM, Lindell D, Chisholm SW (2015) Prochlorococcus: the structure and function of collective diversity. Nat Rev Microbiol 13: 13-27.
109. Nei M, Nozawa M (2011) Roles of mutation and selection in speciation: from Hugo de Vries to the modern genomic era. Genome Biol Evol 3: 812-829.
110. Buchanan M (2015) Bacterial complexity. Nat Phys 11: 887.
111. Fisher JF, Mobashery S (2016) Endless resistance. Endless antibiotics? Med Chem Commun 7: 37-49.
112. Rohwer F, Segall AM (2015) In retrospect: A century of phage lessons. Nature 528: 46-48.
113. Szybalski W, Bryson V (1952) Genetic studies on microbial cross resistance to toxic agents. I. Cross resistance of Escherichia coli to fifteen antibiotics. J Bacteriol 64: 489-499.
114. Toprak E, Veres A, Michel JB, Chait R, Hartl DL, et al. (2012) Evolutionary paths to antibiotic resistance under dynamically sustained drug selection. Nat Genet 44: 101-105.
115. Lazar V, Pal Singh G, Spohn R, Nagy I, Horvath B, et al. (2013) Bacterial evolution of antibiotic hypersensitivity. Mol Syst Biol 9: 700.
116. Baym M, Stone LK, Kishony R (2016) Multidrug evolutionary strategies to reverse antibiotic resistance. Science 351: aad3292.
117. Kolter R, van Wezel GP (2016) Goodbye to brute force in antibiotic discovery? Nat Microbiol 1: 15020.
118. Piskounova E, Agathocleous M, Murphy MM, Hu Z, Huddlestun SE, et al. (2015) Oxidative stress inhibits distant metastasis by human melanoma cells. Nature 527: 186-191.
119. Ling S, Hu Z, Yang Z, Yang F, Li Y, et al. (2015) Extremely high genetic diversity in a single tumor points to prevalence of non-Darwinian cell evolution. Proc Natl Acad Sci U S A 112: E6496-6505.
120. Nikiforov YE, Seethala RR, Tallini G, et al. (2016) Nomenclature revision for encapsulated follicular variant of papillary thyroid carcinoma: A paradigm shift to reduce overtreatment of indolent tumors. JAMA Oncol.
121. Hoption Cann SA, van Netten JP, van Netten C, Glover DW (2002) Spontaneous regression: a hidden treasure buried in time. Med Hypotheses 58: 115-119.
122. Brodeur GM, Bagatell R (2014) Mechanisms of neuroblastoma regression. Nat Rev Clin Oncol 11: 704-713.

123. Diede SJ (2014) Spontaneous regression of metastatic cancer: learning from neuroblastoma. Nat Rev Cancer 14: 71-72.
124. Sekikawa A, Horiuchi BY, Edmundowicz D, Ueshima H, Curb JD, et al. (2003) A "natural experiment" in cardiovascular epidemiology in the early 21st century. Heart 89: 255-257.
125. Feder ME, Mitchell-Olds T (2003) Evolutionary and ecological functional genomics. Nat Rev Genet 4: 651-657.
126. Louis EJ (2007) Evolutionary genetics: making the most of redundancy. Nature 449: 673-674.
127. Hosios AM, Hecht VC, Danai LV, Johnson MO, Rathmell JC, et al. (2016) Amino acids rather than glucose account for the majority of cell mass in proliferating mammalian cells. Dev Cell 36: 540-549.
128. Tanner JM, Rutter J (2016) You are what you eat... or are you? Dev Cell 36: 483-485.
129. Wong S, Slavcev RA (2015) Treating cancer with infection: a review on bacterial cancer therapy. Lett Appl Microbiol 61: 107-112.
130. Ledford H (2016) Cocktails for cancer with a measure of immunotherapy. Nature 532: 162-164.
131. Willyard C (2016) Cancer therapy: an evolved approach. Nature 532: 166-168.
132. Laplane L (2016) Cancer Stem Cells: Philosophy and Therapies: Harvard University Press.
133. Ehsani S (2013) Correlative-causative structures and the 'pericause': an analysis of causation and a model based on cellular biology. arXiv:13100507.
134. Karmon A, Pilpel Y (2016) Biological causal links on physiological and evolutionary time scales. Elife 5: e14424.
135. Oster E (2015) Unobservable Selection and Coefficient Stability: Theory and Validation. Brown University.
136. Hilfinger A, Norman TM, Paulsson J (2016) Exploiting Natural Fluctuations to Identify Kinetic Mechanisms in Sparsely Characterized Systems. Cell Syst 2: 251-259.
137. Elf J (2016) Staying Clear of the Dragons. Cell Syst 2: 219-220.
138. Justman Q (2016) The Power of Logic and Reason. Cell Syst 2: 215.
139. Delling M, Indzhykulian AA, Liu X, Li Y, Xie T, et al. (2016) Primary cilia are not calcium-responsive mechanosensors. Nature 531: 656-660.
140. Norris DP, Jackson PK (2016) Cell biology: Calcium contradictions in cilia. Nature 531: 582-583.
141. Scholl R, Nickelsen K (2015) Discovery of causal mechanisms: Oxidative phosphorylation and the Calvin-Benson cycle. HPLS 37: 180-209.
142. Wallach D, Kang TB, Dillon CP, Green DR (2016) Programmed necrosis in inflammation: Toward identification of the effector molecules. Science 352: aaf2154.
143. Anders HJ, Jayne DR, Rovin BH (2016) Hurdles to the introduction of new therapies for immune-mediated kidney diseases. Nat Rev Nephrol 12: 205-216.
144. Lieshout P, Ben-David B, Lipski M, Namasivayam A (2014) The impact of threat and cognitive stress on speech motor control in people who stutter. J Fluency Disord 40: 93-109.
145. Chen R, Shi L, Hakenberg J, Naughton B, Sklar P, et al. (2016) Analysis of 589,306 genomes identifies individuals resilient to severe Mendelian childhood diseases. Nat Biotechnol.
146. Steffan JS (2016) A cause for childhood ataxia. Elife 5: e14523.
147. Byrd AL, Segre JA (2016) Infectious disease. Adapting Koch's postulates. Science 351: 224-226.
148. Itzhaki RF, Lathe R, Balin BJ, Ball MJ, Bearer EL, et al. (2016) Microbes and Alzheimer's Disease. J Alzheimers Dis 51: 979-984.
149. De Strooper B, Karran E (2016) The Cellular Phase of Alzheimer's Disease. Cell 164: 603-615.
150. Losick R, Desplan C (2008) Stochasticity and cell fate. Science 320: 65-68.
151. Capp JP (2012) Stochastic gene expression stabilization as a new therapeutic strategy for cancer. Bioessays 34: 170-173.
152. Kadelka C, Murrugarra D, Laubenbacher R (2013) Stabilizing gene regulatory networks through feedforward loops. Chaos 23: 025107.
153. Uphoff S, Lord ND, Okumus B, Potvin-Trottier L, Sherratt DJ, et al. (2016) Stochastic activation of a DNA damage response causes cell-to-cell mutation rate variation. Science 351: 1094-1097.
154. Ehsani S (2012) Simple variation of the logistic map as a model to invoke questions on cellular protein trafficking. arXiv:12065557.
155. Kryazhimskiy S, Rice DP, Jerison ER, Desai MM (2014) Microbial evolution. Global epistasis makes adaptation predictable despite sequence-level stochasticity. Science 344: 1519-1522.
156. Wasserstein RL, Lazar NA (2016) The ASA's statement on p-values: context, process, and purpose. Am Stat.
157. Kannan S, Tse D (2015) Fundamental limits of search. Cell Syst 1: 102-103.
158. Markov IL (2014) Limits on fundamental limits to computation. Nature 512: 147-154.
159. Walsh CT (2015) A chemocentric view of the natural product inventory. Nat Chem Biol 11: 620-624.
160. Gregory R (2000) Reversing Rorschach. Nature 404: 19.
161. Casebeer WD (2003) Moral cognition and its neural constituents. Nat Rev Neurosci 4: 840-846.
162. Greene J (2003) From neural 'is' to moral 'ought': what are the moral implications of neuroscientific moral psychology? Nat Rev Neurosci 4: 846-849.
163. Fleming SM (2016) Changing our minds about changes of mind. Elife 5: e14790.
164. Sammut M, Cook SJ, Nguyen KC, Felton T, Hall DH, et al. (2015) Glia-derived neurons are required for sex-specific learning in C. elegans. Nature 526: 385-390.
165. Markram H, Muller E, Ramaswamy S, Reimann MW, Abdellah M, et al. (2015) Reconstruction and Simulation of Neocortical Microcircuitry. Cell 163: 456-492.

166. Chatterjee R (2015) Out of the darkness. Science 350: 372-375.
167. Bizzi E, Ajemian R (2015) A hard scientific quest: Understanding voluntary movements. Daedalus 144: 83-95.
168. (2015) Estimating the reproducibility of psychological science. Science 349: aac4716.
169. Deutsch D (2015) The logic of experimental tests, particularly of Everettian quantum theory. arXiv:150802048.
170. Abbott BP, Abbott R, Abbott TD, Abernathy MR, Acernese F, et al. (2016) Observation of Gravitational Waves from a Binary Black Hole Merger. Phys Rev Lett 116: 061102.
171. Abreu D, Pearce D, Stacchetti E (2012) One-sided uncertainty and delay in reputational bargaining. Economic Theory Center Working Paper No 45-2012.
172. Hauser MD (2016) Challenges to the What, When, and Why. Biolinguistics 10: 1-5.
173. Oates AC, Gorfinkiel N, Gonzalez-Gaitan M, Heisenberg CP (2009) Quantitative approaches in developmental biology. Nat Rev Genet 10: 517-530.
174. Chomsky N (2014) The Minimalist Program (20th Anniversary Edition). MIT Press. pp. vii-xiii.

**Figure 1**

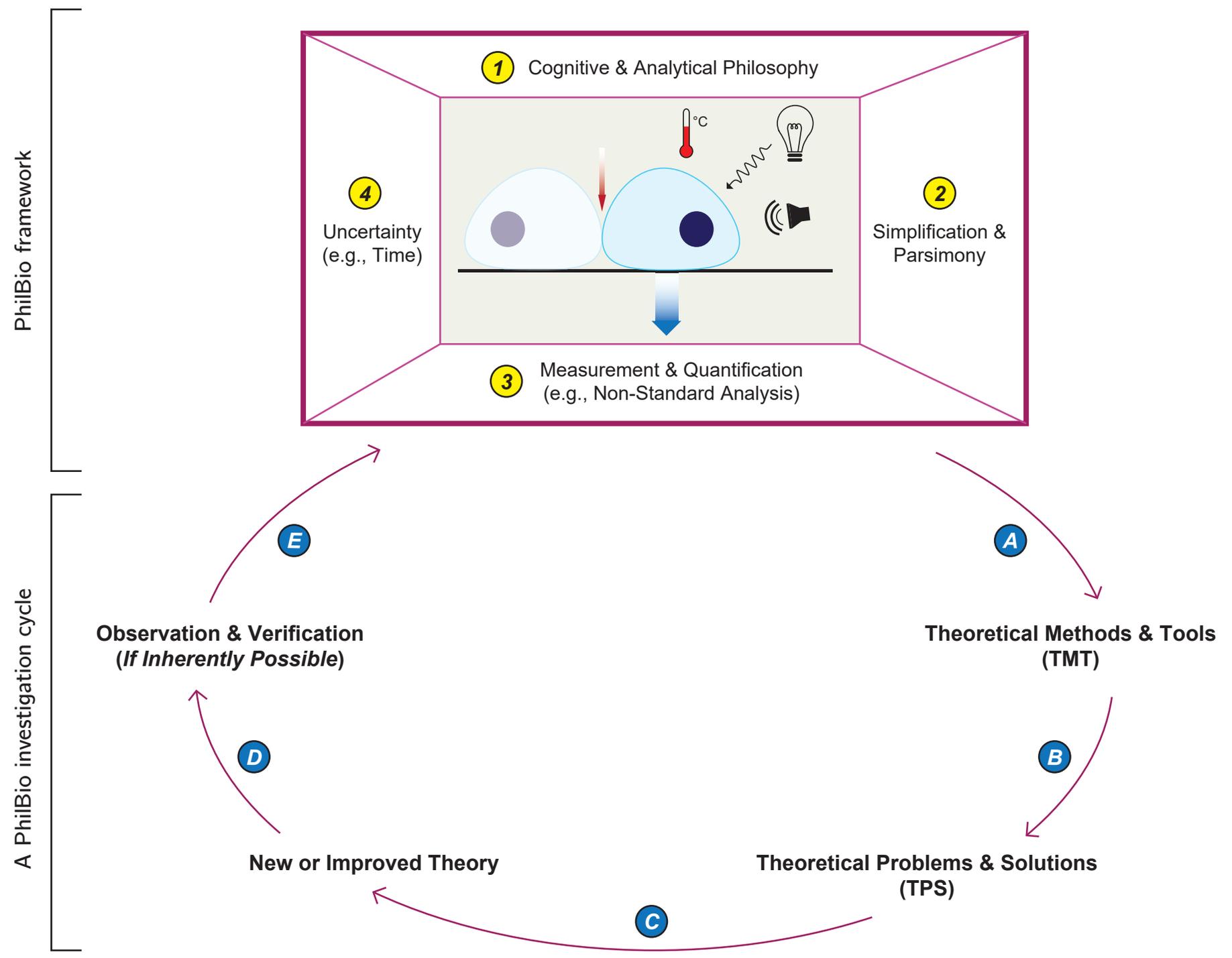